\documentclass{elsart}
\usepackage[]{epsfig}
\usepackage{pstricks}
\usepackage{color,fancyhdr,graphicx,amsmath}
\begin{document}
 
\begin{frontmatter}

 \title{Maxwell's equal area law and the Hawking-Page phase transition.}
 \author{Euro Spallucci\thanksref{infn}}
\thanks[infn]{e-mail address: spallucci@ts.infn.it }
\address{Dipartimento di Fisica Teorica, Universit\`a di Trieste
and INFN, Sezione di Trieste, Italy}
 
\author{Anais Smailagic\thanksref{infn2}}
\thanks[infn2]{e-mail address: anais@ts.infn.it }
\address{Dipartimento di Fisica Teorica, Universit\`a di Trieste
and INFN, Sezione di Trieste, Italy}
       
 \begin{abstract}
In this paper  we study the phases of a Schwarzschild black hole in the Anti deSitter background geometry.
Exploiting fluid/gravity duality we construct the Maxwell equal area \emph{isotherm} $T=T^\ast$
in the temperature-entropy plane, in order to eliminate negative heat capacity BHs. The construction
we present here is reminiscent of the isobar cut in the pressure-volume plane which eliminates unphysical part
of the Van der Walls curves below the critical temperature.
Our construction also modifies the Hawking-Page phase transition. Stable BHs are formed at the temperature $T > T^\ast $, 
while pure radiation persists for $T< T^\ast$.
$T^\ast$ turns out to be below the standard Hawking-Page temperature and there are no unstable BHs as in the usual
scenario. Also, we show that in order to reproduce the correct BH entropy $S=A/4$, one has to write
a black hole equation of state, i.e. $P=P(V)$, in terms of the geometrical volume $V=4\pi r_H^3/3$. 
 \end{abstract}
 \end{frontmatter}
 
\section{Introduction}

Black holes (BHs) are among the most intriguing solutions of Einstein equations. Their geometric description
is fully provided by the theory of General Relativity and is discussed in many excellent textbooks. But, this
is only half of the story.\\
Since the original works by Bekenstein and Hawking some new aspects of the BH behavior emerged once quantum field theory
is coupled to a BH background geometry.  Even if this is only a ``semi-classical'' quantum gravity formulation, the
outcome has profoundly changed the prospective of the BH behavior.\\
A stellar mass, \emph{classical}, BH is characterized by the unique feature of  being a perfect absorber with a vanishing
luminosity. From a thermodynamical point of view, a classical BH is a zero temperature black body.\\
However, a nuclear size BH, interacting with quantized matter, are almost perfect black bodies as they emit
black body radiation at a characteristic non-vanishing temperature! Moreover, BHs are assigned a thermodynamical property
identified with entropy. Thus, there are to complementary descriptions of BH physics: one in terms of pure space-time
geometry, and the other in terms of thermodynamics. The two description are related to each other through the so-called
``\textit{first law}'' of BH (thermo)dynamics

\begin{equation}
 dM = T dS + \phi_H dQ + \Omega_H dJ
\label{primalegge}
\end{equation}

\begin{eqnarray}
 && M=\hbox{total mass energy}\ ,\nonumber\\
 && T =\hbox{Hawking temperature}\ ,\nonumber\\
 && S=\hbox{Entropy}\ ,\nonumber\\
 && \phi_H= \hbox{Coulomb potential on the horizon}\ ,\nonumber\\
 && Q =\hbox{Electric charge}\ ,\nonumber\\
 && \Omega_H=\hbox{Angular velocity of the horizon}\ ,\nonumber\\
 && J=\hbox{Angular momentum}\nonumber
\end{eqnarray}

The first law (\ref{primalegge}) is the basis of the thermodynamical description of the BH as a ``~fluid~''
where $dM$ is the variation of total energy split into variation of ``internal'', ``electrostatic'' and ``rotational'' piece,
which are then given a thermodynamical meaning.    
If compared to the first law as it appears in fluid thermodynamics, one cannot but notice the absence of
a fundamental pair of canonical thermodynamical variables, i.e.
the volume and the pressure of the system. While the volume in question can be naturally identified with the volume excluded by the
event horizon \cite{Kastor:2009wy}, it is a bit ambiguous what geometrical quantity is to be identified with 
pressure \cite{Chamblin:1999tk,Chamblin:1999hg,Tsai:2011gv,Nicolini:2011dp}. 
However, for BHs in Anti/deSitter background there is a \emph{natural} candidate, i.e. the cosmological constant 
\cite{Dolan:2010ha,Dolan:2011xt,Dolan:2011jm,Dolan:2012jh}, \cite{Caldarelli:1999xj,Cvetic:2010jb,Lu:2012xu}. 
In fact, AdS/dS spaces can be sourced by an energy-momentum tensor satisfying the vacuum state equation

\begin{equation} 
P=-\rho=-\frac{\Lambda}{8\pi G_N}
\label{lambda}
\end{equation}

With this identification, it is possible to write the extended first-law  in its simplest form  as

\begin{equation}
 dM = TdS +VdP
\label{primaleggext}
\end{equation}
The equation (\ref{primaleggext}) can be rewritten as 

\begin{equation}
 d(M- pV) = TdS -PdV
\label{primalegge2}
\end{equation}

which allows the identification of the proper thermodynamical potential for the BH:

\begin{eqnarray}
 && U= M -PV\ ,\\
 && M\equiv H = U + PV\ ,\\
 && F=U-TS\ ,\\
 && G=H-TS
\end{eqnarray}

where, $U$ is the internal energy, $H$ is the enthalpy, $F$ is the Helmoltz free-energy and $G$ is the Gibbs free-energy.\\ 
Equation (\ref{primaleggext}) envisages the BH as a ``gas'' enclosed in its geometrical volume, $V$, and subject to an external pressure $P$
given in equation (\ref{lambda}). The question naturally arises: what is the equation of state for the gas?\\
The answer is : take the expression of the Hawking temperature in terms of the geometrical quantities, i.e. radius of the event horizon and
$\Lambda$, replace $\lambda$ with $P$ and solve $P= P\left(\, T, V\,\right)$.  The BH fluid equation makes it possible to investigate
the existence of different ``phases'' of the system and the kind of 
changes of state \cite{Banerjee:2010da,Banerjee:2011au,Banerjee:2011raa,Smailagic:2012cu}. \\
Recently,this approach has led to the identification, for charged AdS BH, of the pressure equation with the one of a Van der Waals
gas \cite{Chamblin:1999tk,Chamblin:1999hg,Tsai:2011gv}. Once this identification is achieved, one can pursue the
standard Van der Waals analysis of phase transitions, calculate the critical exponents and apply the Maxwell equal area construction in
the $(P,V)$ plane. At this point, it is worth mentioning that the Maxwell equal area construction can be equally applied
in the $(T,S)$ plane at constant pressure. This has been done in \cite{Spallucci:2013osa} with
interesting results: i) the equal area law can be analytically solved; ii) the unphysical negative specific heat region is cut-off;
iii) a new BH phase structure emerges; iv) the role of the Van der Waals un-shrinkable molecular volume in taken by the extremal 
BH configuration, thus justifying its stability.\\
The Van der Waals description of charged BHs assumes a particular interest in the framework of  AdS/CFT duality which conjectures 
a relation between  QCD and real gas phase transitions.\\
For neutral, non-rotating, BHs in AdS background the liquid-gas co-existence is absent in the sense that below certain temperature
there are no more BHs. However, in this case one speaks of a different type of phase transition between a pure background radiation
and stable/unstable BHs \cite{Hawking:1982dh}.\\ 
 In this paper, we shall apply the Maxwell equal area construction in the $(T,S )$ plane, to the neutral BH in the AdS background,
and study how the Hawking-Page phase transition is affected.
Although the model does not belong to the same universality class of charged AdS BHs , it still has all the  thermodynamical
potentials  necessary for Maxwell equal area law construction.

\section{Neutral AdS black hole and the area law}
        
  \subsection{Review of the Hawking-Page phase transition. }
 
         We start from the
         Schwarzschild Anti-deSitter (SAdS) metric. 
          The line element  is given by ($G_N=1$ units)
          \begin{equation}
  ds^2=-\left(\, 1 -\frac{2M}{r}  +\frac{r^2}{l^2}\,\right) dt^2 +
          \left(\, 1 -\frac{2M}{r}  +\frac{r^2}{l^2}\,\right)^{-1} dr^2 + r^2\, d\Omega^2_{(2)}
 \label{uno}
 \end{equation}
         where, $M$ is the mass and the AdS curvature radius $l$
         is related to the cosmological constant
         as $\Lambda =-3/l^2$.  We briefly review the properties of the solution (\ref{uno})
         which are  relevant for our discussion.\\ 
        The horizon is determined by the zero of the $g_{rr}^{-1}$ metric component

        \begin{equation}
         M= \frac{r_+}{2}\left(\, 1 + \frac{r_+^2}{l^2}\,\right)
        \label{ADM}
        \end{equation}
          
         In the case of Schwarzschild AdS equation (\ref{ADM}) shows that for any positive value of $M$ there exists only
         one horizon. As a consequence, this kind of BH does not admit any extremal configuration where $M$ has a minimum.
         As a function of $r_+$, $M$ is a monotonically increasing function linearly vanishing as $r_+\to 0 $.\\

\begin{figure}[h!]
               \begin{center}
                \includegraphics[width=15cm,angle=0]{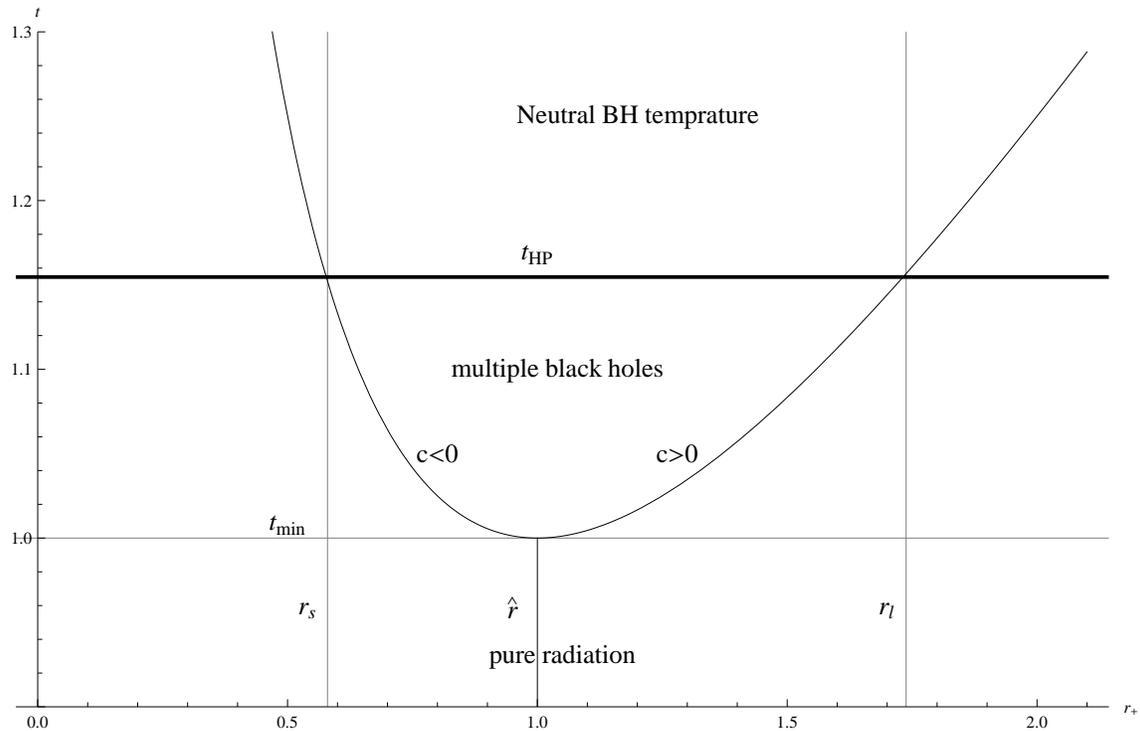}
                \caption{Plot  of the Hawking temperature showing Hawking-Page phases. For $T<T_{min.}$
there are no BHs, but a pure radiation phase. For $T>T_{min}$ there exists a pair of BHs with radii  $r_l,r_s$ given in equation (\ref{radii}).
The smaller BH is (thermodynamically)unstable, while the larger is  either locally stable, $T< T_{HP}$, or globally stable for $T> T_{HP}$. }\label{temp}
                  \end{center}
                 \end{figure}

The corresponding Hawking temperature $T_H$ is given by (see Figure(\ref{temp})
          \begin{equation}
          k_B T = \frac{1}{4\pi \, r_+}\left(\, 1 +\frac{3r^2_+}{l^2}\,\right)
          \label{ht}
           \end{equation}

          The  equation (\ref{ht}) can be inverted by expressing $r_+$ as a function of $T$ as the free variable. Written in
          this form, equation (\ref{ht}) defines the existence of \emph{multiple } BH at the \emph{same} temperature
           
              \begin{equation}
             \frac{3}{l^2} r_+^2 -4\pi k_B T r_+ +1=0
              \end{equation}
               
           This is a quadratic algebraic equation admitting real solutions for
  
         \begin{equation}
          T\ge T_{min.}\equiv \frac{\sqrt{3}}{2\pi l}
         \end{equation}

         For $T< T_{min.}$ there are no BH and we are in a pure radiation phase. The background heat bath is too cold
         to admit nucleation of BHs.\\
         For $T=T_{min}$ a single BH is formed with a radius
   
         \begin{equation}
          \hat{r} = \frac{l}{\sqrt{3}}
         \end{equation}

\begin{figure}[h!]
               \begin{center}
                \includegraphics[width=15cm,angle=0]{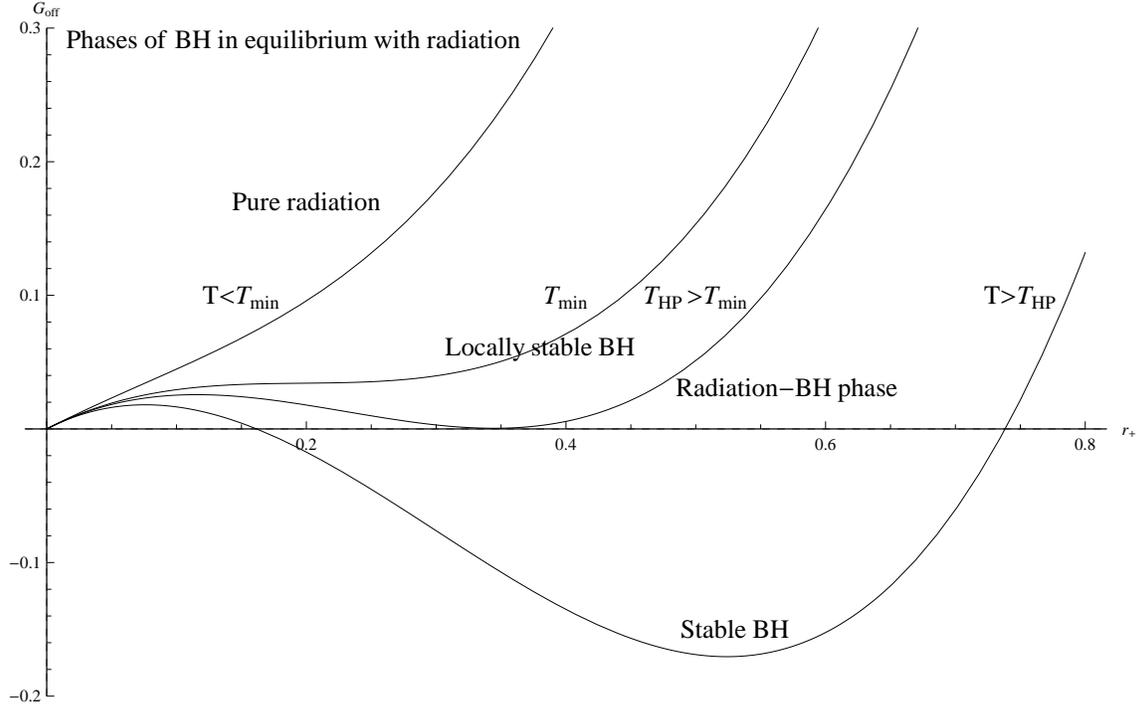}
                \caption{Plot  of  the Gibbs free energy for different temperatures }\label{Gibbs}
                  \end{center}
                 \end{figure}

          For $T>T_{min}$ a pair of BHs (large/small) exists with radii given by

\begin{equation}
r_{l,s} = \frac{T }{2\pi T^2_{min.}}\left(\, 1 \pm\sqrt{1-\frac{T^2_{min.}}{T^2}}\,\right)
\label{radii}
\end{equation}
where $r_s < \hat{r}$ and $r_l > \hat{r}$. 
In order to investigate stability of small and large BHs, we look at the Gibbs free-energy (see Figure(\ref{Gibbs}))

\begin{equation}
 G\equiv  M - TS =\frac{r_+}{2}\left(\, 1 + \frac{r_+^2}{l^2}\,\right)-\pi k_B T r_+^2
\end{equation}

BHs in thermal equilibrium with the background correspond to stationary points of $ G $ 
for $T>T_{min}$. At a lower temperature the single, global, minimum of $G$ is the origin, i.e. $r_+ =0$ and the system is in a pure 
radiation phase.\\
For $T=T_{min}$, $G$ exhibits an \emph{inflexion point} for $r_+=1/2\pi T_{min.}=l/\sqrt{3}$. From this temperature on, we recover two
BHs with radii given by equation (\ref{radii}). The smallest corresponds to a local maximum of $G$, while the larger one is locally
stable being a local minimum. With increasing $T$ the local minimum  of $G$ lowers until reaching zero for
\begin{equation}
 T=\frac{1}{\pi l}\equiv T_{HP}
\end{equation}

$T_{HP}$ is known as the \emph{Hawking-Page critical temperature} where the Gibbs energy of a BH of radius $r_+= l$ is degenerate with
the Gibbs free energy of a pure radiation at the same temperature.
In this case, a BH with radius $r_+=l$ is degenerate with the background radiation. Finally, for $T> T_{HP}$ large BHs are 
globally stable, absolute minima of $G$. \\
 This is a brief summary of the radiation into BH transition as originally introduced by Hawking and Page. \\
 
The question of stability of the BHs is also related to the sign of 
the constant specific heat \cite{Caldarelli:1999xj} 
  
\begin{equation}
   C_P\equiv \left(\, \frac{\partial M}{\partial T}\,\right)_P=-2\pi r_+^2 \left(\, 1+ \frac{3r_+^2}{l^2}\,\right)\left(\,  1- \frac{3r_+^2}{l^2}\,\right)^{-1}
\label{capacity}  
\end{equation}

 $C_P<0$ describes unstable BHs  corresponding to local maxima of $G$, which cannot be in thermal equilibrium with the surrounding heat bath. 
\\
In this paper we shall discuss how to remove negative specific heat BHs by applying Maxwell equal area construction, and see how it affects
the Hawking-Page phase structure. 

    \subsection{The gas equation.}
     Recently,  a number of papers have dealt with the analogy between BHs in AdS and Van der Walls  ``real'' fluids. 
     From this vantage point, a particular attention has been given to charged and/or rotating BHs. This approach is
     made possible if one assign the cosmological constant the role of pressure in the dual VdW picture. The advantage
     of the dual picture is that one can exploit well-known properties and computational techniques from
     the WdV fluid dynamics, to gain
     a deeper understanding of BH thermodynamics, such as critical behavior and phase transitions. Furthermore, it is
     known that phase transitions in real fluids match the experimental results only if one applies an isothermal cut, Maxwell
     equal area law, in the $(P\ , V)$ plane.\\
     By exploiting this kind of duality, we have shown in a recent paper that the equal area construction can be implemented
     in the gravitational sector as well \cite{Spallucci:2013osa}.  
     Working in $\left(\, T, S,\right)$ plane, where $T$ and $S$ are the Hawking temperature and the BH entropy,
     has given a significant advantage in the sense that the area law can be solved analytically and the negative heat capacity
     region can be eliminated. \\
     In this paper we shall extend the same approach to the Schwarzschild AdS case.  While the liquid-gas
     transition in the WdV fluid is dual to a small-large BH change in the gravitational picture, the Hawking-Page transition,
      for neutral BHs,
      has been interpreted by Witten as the gravitational dual of the QCD de-confinement transition \cite{Witten:1998zw,Witten:1998qj}. 
      Thus, the study of the Schwarzschild
      AdS BH properties can shed more light on the quark-gluon plasma properties presently subject to an intensive research
      interest \cite{Myers:2008fv}. \\
      In order to obtain the fluid picture of the SAdS BH we rewrite equation (\ref{ht}) as

      \begin{equation}
           P = \frac{k_B T}{2 r_+ } -\frac{1}{8\pi \, r_+^2}
          \label{prh}
       \end{equation}
  
       where the duality ``dictionary'' called for substitution:

       \begin{equation}
        -\frac{\Lambda}{8\pi}\longrightarrow P
       \end{equation}

    \begin{figure}[h!]
               \begin{center}
                \includegraphics[width=15cm,angle=0]{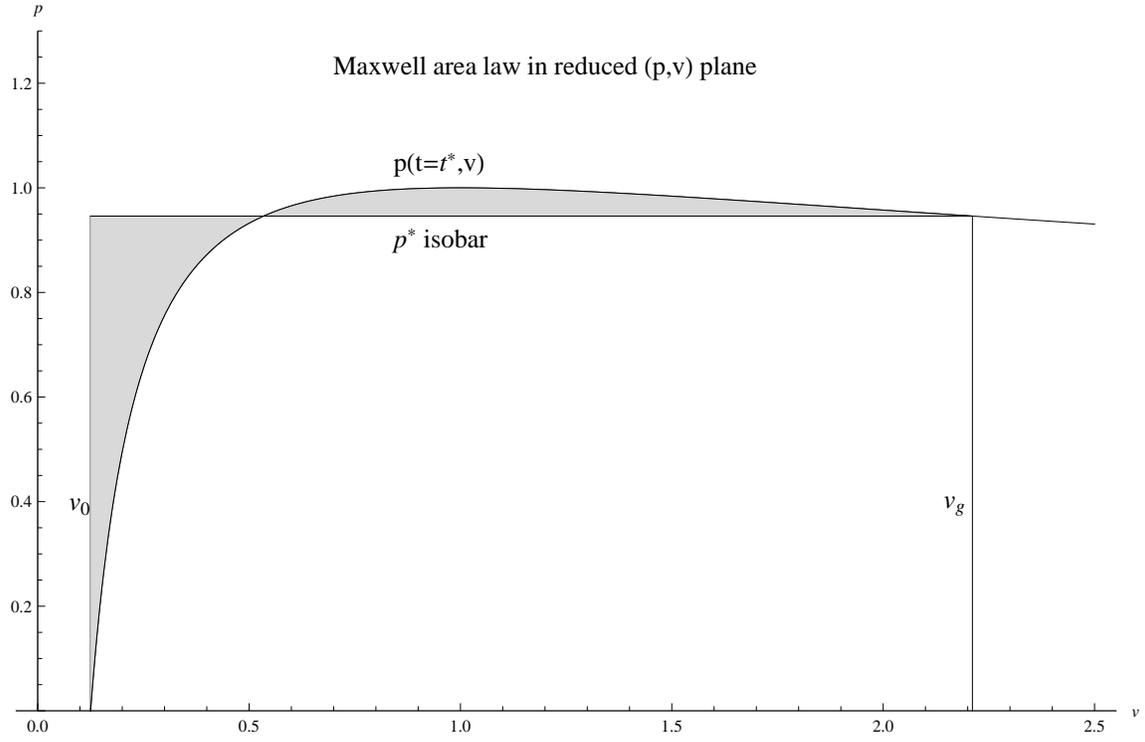}
                \caption{ Plot  of the (rescaled) pressure at fixed $T$, as a function of the geometric volume of the BH. 
                 Equal area regions are colored in black.}\label{pressure}
                  \end{center}
                 \end{figure}

The graph of $P$ for constant $T$ is given in the Figure(\ref{pressure}).  One sees that $P$
 vanishes at $r_0= 1/4\pi k_B T$, and reaches maximum  $P_M= \pi k_B^2 T^2/2G_N$ at $r_M=2r_0 $. \\
Although the graph is not of VdW form, the equal area law can be still defined between  $r_0$ and $r_g$ which
will be determined later. In order to make equation (\ref{prh})  look like the state equation of a fluid, we still need to associate
a proper volume to the system. \\
In previous works \cite{Kubiznak:2012wp,Gunasekaran:2012dq} the specific volume of the fluid was 
identified as $v=2 r_H G_N$. With respect to this choice, we cannot help but notice that: 
\begin{itemize} 
\item
the conjugate canonical variable to the cosmological constant
(identified with the \emph{inward} pressure ) is the geometric volume $V=4\pi r_H^3/3$ of the BH since

\begin{equation}
 V \equiv \frac{\partial H}{\partial P}= \frac{4\pi}{3} r_+^3
\end{equation}
where, $H$ is given by equation (\ref{ADM}).
\item Furthermore, it is widely accepted that the BH entropy is given by the area law $S= \pi r_+^2$.
On the other hand,  the entropy
of a gas is logarithmic in the specific volume $v$.
\end{itemize}

Therefore, in order to satisfy both requirements one has to express equation (\ref{prh}) in terms of the geometric volume as

 \begin{equation}
           P = \frac{k_B T}{2 }\left(\, \frac{4\pi}{3V}\,\right)^{1/3} -\frac{1}{8\pi}\left(\, \frac{4\pi}{3V}\,\right)^{2/3}
          \label{prh2}
           \end{equation}
where $k_B$ is the Boltzman constant.\\
To compute the entropy, we start from the general formula for a gas:

\begin{equation}
 dS =C_V\,\frac{dT}{T}+\left(\frac{\partial P}{\partial T}\right)_V dV
\end{equation}
where $C_V$ is the specific heat at constant volume.  Thus, equation (\ref{prh2}) leads to

\begin{align*}
  & dS =\frac{k_B}{2}\left(\frac{4\pi}{3}\right)^{1/3}\frac{dV}{V^{1/3}}\\
  & S =k_B\left(\frac{A}{4}\right)\\
\end{align*}

We find the BH \emph{area law} for entropy. 
Therefore, the area law follows from the gas equation only in terms of the \textbf{geometric} volume of the BH.

\section{Equal area Maxwell construction.}

The equal area law was introduced by Maxwell in order to explain the experimental behavior of real fluid. Normally, this construction
is applied in the $(P,V)$ plane keeping the temperature constant. Theoretically, it can be derived from the variation
of the Gibbs free energy $G$:

\begin{equation}
 dG = - SdT + VdP
\label{dgibbs}
\end{equation}

At constant $T$ one finds the equal area law:

\begin{equation}
 P^\ast \Delta V= \int_{V_l}^{V_g} PdV
\end{equation}

where, $P=P^\ast$ is an isobar defining equal areas, and $V_l$, $V_g$ are the volume of the liquid and gaseous phase. 
However, if one keeps the pressure constant in (\ref{dgibbs}), the same construction can be done in the $(T,S)$ plane.\\
In this paper,
we construct the Maxwell area law in the latter plane for two reasons:
\begin{itemize}
 \item it is through the temperature graph that the Hawking-Page phase 
transition has been originally studied \cite{Hawking:1982dh}, and we are interested in the modification of this
phase structure.
\item We have shown in \cite{Spallucci:2013osa} that in this plane the equation for the equal area law can be explicitly solved.
\end{itemize}

We start from the temperature as a function of entropy  given by 
 
\begin{equation}
          k_B T = \frac{1}{4\sqrt{\pi \, S}}\left(\, 1 +\frac{3S}{\pi l^2}\,\right)
          \label{ts}
           \end{equation}
\begin{figure}[h!]
               \begin{center}
                \includegraphics[width=15cm,angle=0]{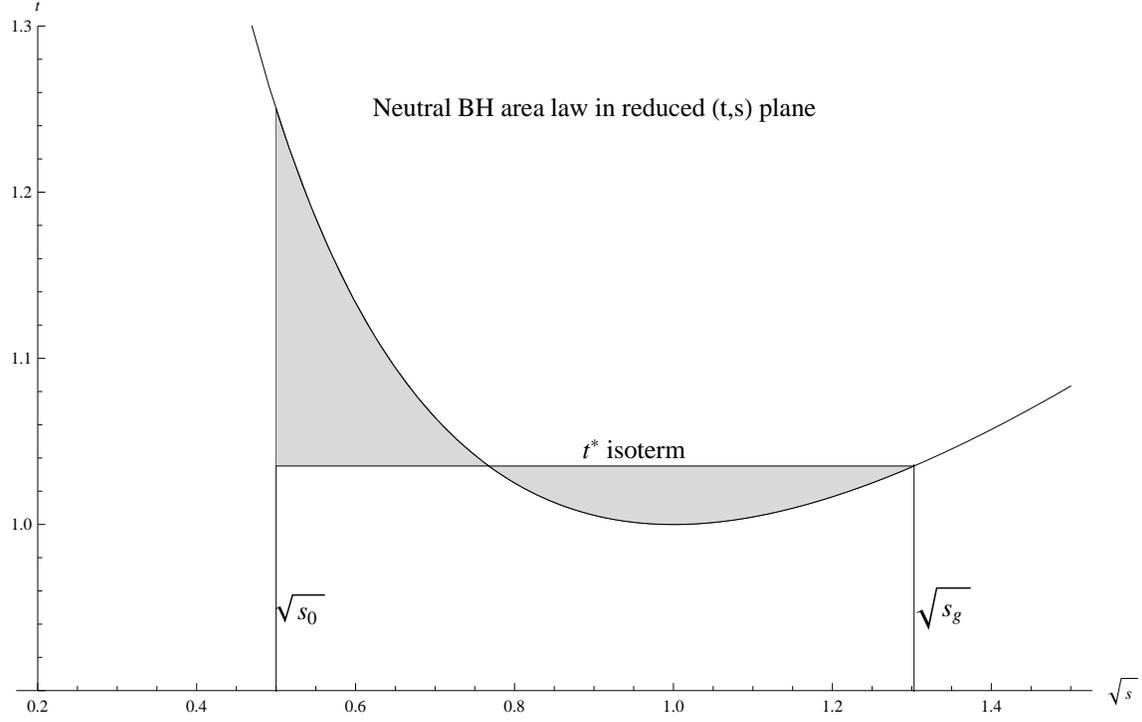}
                \caption{Plot  of the temperature as function of the entropy. Equal area regions area
                shown in black. $s_0$, $s_g$ e $t^ast$ are given in equations (\ref{sol1}), (\ref{sol2}) and (\ref{sol4})}
                \label{TH2}
                  \end{center}
                 \end{figure}
and plotted in Figure(\ref{TH2}). The graph has  a minimum  at $S_{min.}= \pi l^2/3$ and $k_B T_{min}=\sqrt{3}/2\pi l $. 
By rescaling equation (\ref{ts}) with respect to these extremal values we obtain the reduced form 

\begin{align*}
t&=\frac{1}{2}\sqrt{s}+\frac{1}{2}\frac{1}{\sqrt{s}}\\
s&\equiv \frac{S}{S_{min}}\\
t&\equiv \frac{T}{T_{min}}\\
\end{align*}
which is valid for any value of the cosmological constant.
Maxwell area law follows from requirement that the Gibbs free-energy is the same for co-existing BHs.  One finds 

\begin{align*}
	& G=M-T\,S\\
	& \Delta G_{1,2}=-\int_1^2\,SdT=0\\
	& T^*\Delta S_{1,2}=\int_1^2\,T\,dS\\
  \end{align*}	
which leads to the definition of the equal area isotherm 

\begin{equation}
T^*=\frac{1}{2\sqrt{\pi}\left(\sqrt{S_1}+\sqrt{S_2}\right)}\left[1+\frac{8\,P}{3}\left(S_1+S_2+\sqrt{S_1\,S_2}\right)\right]
\end{equation}

Reduced entropy solutions to the equal area law are obtained solving simultaneously the following system 

\begin{align}
t&=\frac{1}{2\sqrt{s}}\left(s+1\right)\\
t^*&=\frac{1}{\left(\sqrt{s_0}+\sqrt{s}\right)}\left[\, 1+\frac{1}{3}\left(s_0+s+\sqrt{s_0\,s}\right)\,\right]
\end{align} 

Introducing $\sqrt{s}\equiv x$ the above system leads to the equation

 \begin{equation}
  2x^3+x^2-7x+3=0
\end{equation}

which has solutions
\begin{align} 
& \sqrt{s_0}= x_1=\frac{1}{2}\label{sol1}\\
& \sqrt{s_g} = x_2=\frac{1}{2}\left(\sqrt{13}-1\right)\label{sol2}\\
&  x_3=-\frac{1}{2}\left(\sqrt{13}+1\right)\label{sol3}\\
&  t^\ast =\frac{1}{6}\left(2\sqrt{13}-1\right)\label{sol4}
\end{align}

 $x_3$ is an un-physical solution leading to negative entropy, and $t^\ast$ is the Maxwell isotherm.\\

 Maxwell construction has allowed us to \emph{eliminate  negative specific heat} part in the temperature graph. 
 The motivation for removing this part of the temperature graph comes from the fact that it has no plausible
physical explanation and does not allow the system to be in equilibrium with the heat bath. In the $(T,S)$ plane
it is the analogue of the unstable (unphysical) part of the VdW graph in the $(P,V)$ plane. 
The flattening of the temperature graph through the isothermal  cut  modifies the Hawking-Page phase transition in the
following way: \\
i) pure radiation phase survives beyond $T_{min.}=\sqrt{3}/2\pi l $ up to the higher temperature, given by the isotherm 
$T^\ast$ :

\begin{equation}
k_B T^\ast= \frac{\sqrt{3}}{12\pi l}\left(\, 2\sqrt{13}-1\,\right)
\end{equation}

ii) For $T=T^\ast $ multiple, but \emph{stable}, BHs form of different entropy. The are no more locally stable BHs
that can decay into radiation as it occurs in Hawking-Page picture in the interval $T_{min}< T < T_{HP}$. \\   

iii) For $T> T^\ast$, there exists single, stable BH with positive heat capacity.\\
 
One may be tempted to ask what kind of statistical ensemble describes the BH thermodynamics ? \\
Hawking has argued that the suitable description is in terms of canonical ensemble with the generating function
to be defined in terms of Helmoltz free-energy $F= M -T S $. This seems to be a suitable picture as long as the cosmological
constant is held fixed. On the other hand, when one decides to work in an extended phase space \cite{Caldarelli:1999xj,Dolan:2011xt},
the same function is interpreted as the Gibbs free-energy and, thus, the suitable description should be in terms
of grand canonical ensemble, since the $PV$ term, a.k.a. as the $\mathbf{q}$-potential, defines the generating function for this ensemble.
Since in this paper we put in evidence the $PV$ term, and allow for $P$ to vary in the $(P,V)$ plane, we are actually dealing
with the grand canonical ensemble. An additional motivation is that the Maxwell area construction is based on the Gibbs free-energy
characterizing the grand canonical ensemble. \\
In our view, a statistical description can be only formal, until a microscopic structure of the BH is properly understood
in terms of a reliable theory of quantum gravity. Nevertheless, working with different ensembles is just a matter of
convenience leading to the same equation of state (\ref{prh2}).


\section{Conclusions }
 
In this paper we have presented a new scenario for Scharzschild BH nucleation in the AdS vacuum.
This kind of BHs have a well known, but largely ignored, problem concerning the negative heat
for  $r_+ < r_{min.}=l/\sqrt{3}$.  Similar negative
specific heat regions exist also for other type of BHs. So far, no plausible physical explanation
for negative heat capacity has been given. The problem of negative heat capacity is 
related to the fact that such BHs cannot be in thermal equilibrium
with the background, thus they cannot slowly evolve through a series of successive equilibrium 
states at different temperature.
To avoid these problems, we construct an isothermal cut in the temperature graph which eliminates negative heat capacity region, 
following Maxwell equal area prescription in the $(T,S)$ plane.
\\
We also studied the effect of the Maxwell construction on the Hawking-Page transition picture. It turns out that there are 
significant modifications: BHs start nucleating at a temperature  $T^\ast$ higher than the temperature 
 $T_{min.}=\sqrt{3}/2\pi l$ as in the original Hawking-Page scenario.  
Above $T^\ast$ there are only single stable BHs. Therefore, having removed the negative specific heat region, 
unstable BHs cannot be formed. Along the isotherm $T=T^\ast $ BHs of different radius have the same
negative free-energy and from a statistical point of view are equally probable among themselves, but more stable than the pure radiation.
Degeneracy between radiation and BHs has been removed since $T^\ast > T_{HP}$.
\\
We have also discussed the corresponding equation of state $P=P(V)$, where the pressure is related to the cosmological constant and
$V$ corresponds to the geometric volume of the BH. In the literature different authors make different choices 
for the volume to be used in the equation of state, and it seems that there is no unique choice
that everybody agrees upon. We have given our contribution to this controversy based on the following argument: 
if a gas equation is to truly describe a BH is has to reproduce its basic properties and, in particular, area/entropy 
relation should follow from thermodynamical arguments. We have shown by an explicit calculation that
the geometrical volume $V=4\pi r_H^3/3$ is needed  to obtain the correct entropy, at least in the model described in this paper.
\\


\end{document}